\documentclass{phb-proc4-auth}

\usepackage{graphicx}
\usepackage{amssymb}

\begin{document}
\begin{frontmatter}

%----------------------------------------------------------------------
% Specify destination and version number of the manuscript

\journal{SCES '04}

%----------------------------------------------------------------------
% Title of manuscript

\title{Field-induced phase transition in the periodic Anderson model}

%----------------------------------------------------------------------
% List of authors
%
% List each author using a separate \author{} command
%
% If there is more than one author address, add a label to each author
% of the form \author[label]{name}.  This label should be identical to
% the corresponding label provided with the \address command.
%
% e.g. if there are three authors from two institutions in USA and 
% France, you can link them to their respective addresses, using
%
% \author[US]{John Doe}
% \author[US,FR]{Jane Doe}
% \author[FR]{Jean Dupont}
% \address[US]{University of Life, Somewhere, USA}
% \address[FR]{Universite de la Vie, Quelque Part, France}
%
% N.B. Unlike the document class used for abstract submissions, it is
% possible to have the author associated with more than one address,
% as shown in the example above.
%

\author{Takuma Ohashi\corauthref{1}}, 
\author{Akihisa Koga}, 
\author{Sei-ichiro Suga}, and 
\author{Norio Kawakami}

%----------------------------------------------------------------------
% List of addresses
%
% If there is more than one address, list each using a separate 
% \address command using a label to link it to the respective author
% as described above
 
\address{Department of Applied Physics, Osaka University, Suita, Osaka 565-0871, Japan}

%----------------------------------------------------------------------
% Title page footnotes
%
% If you need to add qualifying information to any of the authors, 
% use the \thanksref{} command within the \author command.  The 
% argument is the label of a corresponding \thanks[label]{text}
% command which contains the footnote text
%
% e.g. you can acknowledge a funding authority for John Doe, using
%
% \author{John Doe\thanksref{ABC}}
% \thanks[ABC]{This work was supported by Institute of Unphysical 
%    Phenomena under contract no. ABC-123}
%

%\thanks[]{Takuma Ohashi}

%----------------------------------------------------------------------
% Contact Information
%
% Add the complete postal address, telephone number, fax number, and
% email address of the corresponding author as a special footnote using
% the \corauth[]{} command.  This works in a similar way to the \thanks 
% command.  Add the \corauthref{} command within the \author command.
% The argument is the label of a corresponding \corauth[label]{text}
% command which contains the contact information.  Prefix the text with
% Corresponding Author:
%
% e.g. if the contact author is John Doe,
%
% \author{John Doe\corauthref{1}}
% \corauth[1]{Corresponding Author: University of Life, 123 Some St.,
%    Somewhere, MI 12345, USA.  Phone: (555) 555-5555 
%    Fax: (555) 555-7777, Email: JDoe@uol.edu}
%

\corauth[1]{Department of Applied Physics, Osaka University, Yamada-oka 2-1, Suita, Osaka 565-0871, Japan, Email: ohashi@tp.ap.eng.osaka-u.ac.jp}

%----------------------------------------------------------------------
% Text of abstract

\begin{abstract}
We investigate the effect of magnetic fields on a Kondo insulator by 
using the periodic Anderson model. 
The analysis by dynamical mean field theory combined with 
quantum Monte Carlo simulations reveals that the magnetic field 
drives the Kondo insulator to a transverse antiferromagnetic 
insulator at low temperatures. 
We calculate the staggered spin susceptibility and 
find its divergence signaling the antiferromagnetic
instability. 
Further investigation of the spin correlation functions and 
the magnetization process  clarifies how the magnetic field 
suppresses the Kondo singlet formation and induces the 
transverse antiferromagnetic ordering. 
\end{abstract}

%----------------------------------------------------------------------
% Manuscript keywords
%
% Please give two or three keywords in the form: keyword \sep keyword
% e.g. NMR \sep superconductivity
%
% NB The syntax is different from the abstract document class


%----------------------------------------------------------------------
% End of front page

\end{frontmatter}

%----------------------------------------------------------------------
% Manuscript text
%
% Fill in the following space with the manuscript text.
%
% A number of LaTeX commands may be invoked in this space, e.g.
%
% \section{} : to insert a new section title
% \label{}   : to label the numbered section for use in \ref{}
% \cite{}    : to add a reference using the label in \bibitem{}
% 
% A number of LaTeX environments may be used, e.g. 
% \begin{equation}
%     An equation inserted here will be automatically numbered
% \end{equation}  
%
% Please refer to other LaTeX documentation for help on using these
% environments.

Heavy-fermion systems with various ground states have attracted 
continued interest. One of the interesting examples is the Kondo 
insulator (KI), where a small charge gap renormalized by the 
Coulomb interaction appears at low temperatures\cite{ki}.
It is known that this insulating phase gets unstable
upon introducing the magnetic field, the pressure, etc. 
Experiments on 
some KI in high magnetic fields indicate 
closure of the Kondo-insulating gap\cite{field}, 
exemplifying a transition from the KI to a 
correlated metal\cite{saso,mutou}.
If the KI is in the proximity of magnetic instability, 
 the local singlet formation gets weak, and 
an applied field may possibly trigger 
a phase transition to the antiferromagnetic (AF)
ordered state before it becomes a metal.

The periodic Anderson model (PAM)  at half filling may be 
a simplified model to describe  the Kondo insulating 
phase and the AF  phase. 
The magnetic instability of the PAM has been investigated
by a variety of methods\cite{pam_phase1,pam_phase2,jarrell,rozenberg}.
Recently, the magnetic-field effects on the two-dimensional
Kondo lattice model have been studied on the basis of
the mean field theory and quantum Monte Carlo (QMC) 
simulations. It has been 
 found that the magnetic field induces 
a second-order phase transition from the paramagnetic 
to the AF ground state\cite{beach,milat}. 

In this paper, we investigate field-induced AF phase transitions
in the PAM by using  dynamical mean field theory (DMFT) \cite{dmft}
combined with QMC simulations \cite{qmc} at finite temperatures. 
The model we study here is the PAM with the Zeeman splitting, 
%%%%%%%%%%%%%%%%%%%%
\begin{eqnarray}
H &=&
  -t \sum_{\left < i,j \right >,\sigma} c_{i\sigma}^\dag c_{j\sigma}
  + V \sum_{i,\sigma}
  \left [
    c_{i \sigma}^\dag f_{i \sigma} 
    + \mathrm{h.c.}
  \right ] \nonumber \\
  &+& U \sum_i \left [ n^f_{i \uparrow} - 1/2 \right ]
  \left [ n^f_{i \downarrow} - 1/2 \right ] \nonumber \\
  &-& g \mu _B B \sum _i 
  \left [ S^{f}_{i,z} + S^{c}_{i,z} \right ], \nonumber
\end{eqnarray}
%%%%%%%%%%%%%%%%%%%%%%%%%%%%
where all the symbols have their usual meaning. 
We consider the  hyper-cubic lattice with a 
bipartite property in infinite dimensions.
The bare density of states for conduction ($c$) 
electrons is Gaussian 
with the width $t^*$: $\rho_0(\varepsilon)
=\exp[-(\varepsilon / t^*)^2]/\sqrt{\pi t^{*2}}$.
We choose $t^*=1$ as the energy unit. 
The magnetic field $B$ applied along the $z$ direction is
coupled to both of $c$ and $f$ electrons. 
We set $g \mu_B = 1$ for both $c$ and $f$ electrons.

DMFT is a powerful framework to study 
strongly correlated electron systems.
In DMFT, the original lattice model is mapped onto an effective 
impurity model with the self-consistently determined medium, 
which is solved by means of QMC simulations in our approach. 
We use the typical parameters 
$U=2.0$ and $V=0.6$ in the following calculation. 
For these parameters, the ground state at zero field 
is a paramagnetic KI, as shown 
by Jarrell {\it et. al.} \cite{jarrell}. 

 We first show that the magnetic field induces a KI-AF 
phase transition. 
We calculate the staggered spin susceptibility 
$\chi_{xx} ({\bf q}={\bf Q})$, with
${\bf Q} = \left [ \pi,\pi,\ldots \right ] $, 
where the suffix $x$ denotes the direction perpendicular 
to the field. 
In Fig. 1, we plot the inverse of the total susceptibility 
$\chi^{tot}_{xx} ({\bf Q})$ 
and the susceptibility for 
$f$ electrons 
$\chi^{f}_{xx} ({\bf Q})$ as a function of the magnetic field 
for different temperatures. 
At temperatures $T=1/24$ and $T=1/28$, the staggered susceptibilities 
take a maximum at the magnetic field $B \sim 0.28$. 
At the lowest temperature $T=1/32$ the susceptibilities 
diverge at the field $0.25 < B < 0.34$. 
The magnetic field triggers a phase transition from the 
paramagnetic KI to AF ordered state. 
We determine two critical values of the field 
by extrapolating the inverse susceptibility to zero 
with the form 
$\chi _{xx}({\bf Q})^{-1} \propto B-B_c$.

%%%%%%%%%%%%%%%%%%%%%%%%%%%%%%%%%%%%%
\begin{figure}[bt]
\begin{center}
\includegraphics[clip,trim=2.2cm 6.2cm 0.20cm 7cm,width=5cm]{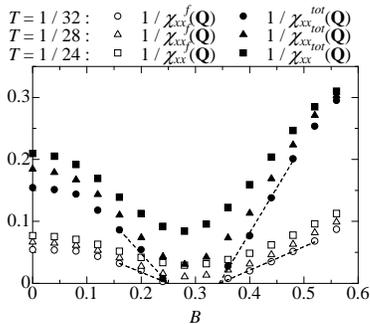}
\end{center}
\caption{%
The inverse of the staggered spin susceptibilities. 
The dashed lines are fitting functions with the form
$\chi _{xx}({\bf Q})^{-1} \propto B-B_c$. 
}
\label{fig1}
\end{figure}
%%%%%%%%%%%%%%%%%%%%%%%%%%%%%%%%%%%%%

 We further calculate the magnetization process and 
the field dependence of the spin correlation functions at 
$T=1/30$, which is slightly higher 
than AF transition temperature. 
The magnetization $M$, defined as
$M = \langle [ n_\uparrow^c + n_\uparrow^f ]
   - [ n_\downarrow^c + n_\downarrow^f ] \rangle$, 
is shown in the top panel of Fig. 2. 
We can see the crossover behavior from the KI with the spin gap to 
the paramagnetic metal around $B \sim 0.28$. 
Around this field, we also see the change in 
the character of the spin correlation functions. 
In the middle of Fig. 2, we show  the variance of the $f$-moment 
in the $z$ direction $\langle [ :S_z^f: ]^2 $, 
where $::$ denotes the normal order, 
and in the $x$ direction $\langle [ S_x^f ]^2 \rangle$. 
When the magnetic field larger than $B \sim 0.28$ is applied, 
spin fluctuations in the $z$ direction are suppressed 
and the moment in the $x$ direction get large, 
since the Kondo singlet formation is suppressed
by a magnetic field. 
The consistent behavior is seen in the spin correlation functions
between the $c$ and $f$ electrons, 
$\left \langle :S_z^c: :S_z^f: \right \rangle$ and 
$\left \langle S_x^c S_x^f \right \rangle$, 
plotted in the bottom of Fig. 2. 
AF correlations between the $c$ and $f$ electrons 
are suppressed by a magnetic field. 
We can see that the correlations in the $z$ direction are more suppressed 
than those in the $x$ direction. 
The above behavior explains why the transverse rather than 
longitudinal antiferromagnetic ordering is favored 
in finite magnetic fields. 

The authors thank T. Saso for valuable discussions. 
A part of computations was done at the Supercomputer Center 
at the Institute for Solid State Physics, University of Tokyo. 
This work was partly supported by a Grant-in-Aid from the Ministry 
of Education, Science, Sports and Culture of Japan. 

%%%%%%%%%%%%%%%%%%%%%%%%%%%%%%%%%%%%%
\begin{figure}[bt]
\begin{center}
\includegraphics[clip,trim=3.5cm 6.2cm 4.2cm 5cm,width=4cm]{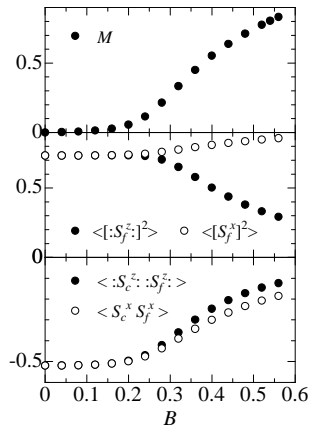}
\end{center}
\caption{%
The magnetization process (top), the field dependence of 
the variance of the spin moment (middle) and 
the spin correlation functions between the $c$ and 
the $f$ electrons (bottom). 
}
\label{fig2}
\end{figure}
%%%%%%%%%%%%%%%%%%%%%%%%%%%%%%%%%%%%%

%----------------------------------------------------------------------
% Reference section
%
% List each reference with a separate \bibitem{} command.  The
% argument contains the label that is used in the \cite{} command
% in the main text
%
% e.g.
%
%    This follows our pioneering work on TdB2\cite{TdB2}.
%
% \bibitem{TdB2}
% J. Doe, J. Doe, and J. Dupont, J. Irrep. Res. 10 (2000) 1000.

%----------------------------------------------------------------------
% Figures and Tables
%
% Insert figures and tables at the end of the document unless you
% are familiar with the LaTeX positional options.
%
% \begin{figure}
%     \centering
%     \includegraphics{filename.eps}
%     \caption{Insert figure caption here} 
% \end{figure}  
%
% \begin{table}
%     \centering
%     \begin{tabular}
%     Insert table here
%     \end{tabular}
%     \caption{Insert table caption here}
% \end{table}  
%
% Please refer to other LaTeX documentation for help on using these
% environments.

%----------------------------------------------------------------------
% Terminate document

\end{document}